\documentclass[prb,aps,twocolumn,showpacs]{revtex4-2} 
\usepackage{amsmath,amssymb,amsthm}
\usepackage{graphicx}
\usepackage{subfigure}
\usepackage{amsmath}
\usepackage{amsfonts,amssymb}
\usepackage{bm}
\usepackage{float}
\usepackage{color}
\usepackage{sidecap}
\allowdisplaybreaks
\usepackage{soul}
\setstcolor{red}
\usepackage[normalem]{ulem}
\usepackage{hyperref}
\hypersetup{colorlinks=true,breaklinks,urlcolor=blue,linkcolor=blue,citecolor=blue}

\begin{document}

\title{Wannier-Stark localization in one-dimensional amplitude-chirped lattices}
\author{Qi-Bo Zeng}
\email{zengqibo@cnu.edu.cn}
\affiliation{Department of Physics, Capital Normal University, Beijing 100048, China}

\author{Bo Hou}
\affiliation{Department of Physics, Capital Normal University, Beijing 100048, China}

\author{Han Xiao}
\affiliation{Department of Physics, Capital Normal University, Beijing 100048, China}

\begin{abstract}
We study the Wannier-Stark (WS) localization in one-dimensional amplitude-chirped lattices with the $j$th onsite potential modulated by a function $Fj\cos(2\pi \alpha j)$, where $F$ is the external field with a period determined by $\alpha=p/q$ ($p$ and $q$ are coprime integers). In the Hermitian (or non-Hermitian) systems with real (or imaginary) fields, we can obtain real (or imaginary) WS ladders in the eigenenergy spectrum. In most cases with $q \geq 2$, there are multiple WS ladders with all the eigenstates localized in the strong field limit. However, in the lattices with $q=4$, the energy-dependent localization phenomenon emerges due to the presence of both spatially periodic and linearly increasing behaviors in the onsite potential. About half the number of eigenstates are gathered at the band center and can extend over a wide region or even the full range of the lattice, even when the field becomes very strong. Moreover, in the non-Hermitian lattices with odd $q$, some of the WS ladders become doubly degenerate, where the eigenstates are evenly distributed at two neighboring sites in a wide regime of field strength. Our work opens an avenue for exploring WS localization in both Hermitian and non-Hermitian amplitude-chirped lattices.
\end{abstract}
\maketitle
\date{today}

\section{Introduction}
Anderson localization~\cite{Anderson1958} is a ubiquitous phenomenon in disordered physical systems~\cite{Lee1985RMP,Evers2008RMP}, which has been experimentally observed in various platforms such as cold atoms~\cite{Billy2008Nat,Roati2008Nat,Luschen2018PRL}, light~\cite{Wiersma1997Nat,Scheffold1999Nat,Schwartz2007Nat}, microwave~\cite{Dalichaouch1991Nat,Pradhan2000PRL}, and photonic lattices~\cite{Lahini2009PRL}. The localization phase transition occurs with a mobility edge in three-dimensional systems~\cite{Mott1987JPhy} but is excluded by scaling theory in lower-dimensional systems with random disorders~\cite{Abrahams1979PRL}. However, it still happens in systems with correlated disorders, such as the famous quasiperiodic Aubry-Andr\'e model~\cite{Aubry1980}. In addition, localization can also emerge in disorder-free systems, which traces back to the Wannier-Stark (WS) lattice subjected to a constant electric field~\cite{Wannier1962RMP,Fukuyama1973PRB,Emin1987PRB,Holthaus1996Phio,Hartmann2004NJP}. The eigenenergies in such systems are equally spaced, forming the WS ladders with all eigenstates becoming exponentially localized in the strong field limit. Recently, such Stark lattices have also been exploited to study the many-body localization phenomenon when further taking the particle interaction into consideration~\cite{Schulz2019PRL,Taylor2020PRB,Guo2021PRL,Morong2021Nat}. So far, most studies on WS localization focus on systems with uniform fields represented by a linearly site-dependent potential in the Hamiltonian. If we further introduce spatially periodic factors and construct the amplitude-chirped lattices~\cite{Leder2016NatCom}, what will happen to the WS ladder and localization properties remains unexplored.

On the other hand, non-Hermitian physics has become a research field undergoing explosive development during the past few years~\cite{Cao2015RMP,Konotop2016RMP,Ganainy2018NatPhy,Ashida2020AiP,Bergholtz2021RMP}. A variety of exotic phenomena, such as the real spectra in $\mathcal{PT}$-symmetric systems~\cite{Bender1998PRL,Bender2002PRL,Bender2007RPP} and the non-Hermitian skin effect (NHSE)~\cite{Yao2018PRL1,Yao2018PRL2}, have been revealed. The existence of non-Hermiticity can modify the band topology~\cite{Bergholtz2021RMP,Yao2018PRL1,Yao2018PRL2,Gong2018PRX} as well as the Anderson localization transition~\cite{Shnerb1998PRL,Jiang2019PRB,Zeng2020PRR,Liu2021PRB1,Liu2021PRB2} in a significant way. Recently, non-Hermitian systems with a uniform field have been studied, where the continuum of bound states appears when the imaginary field is weak~\cite{Wang2023PRL}. When the field gets stronger, imaginary WS ladders will appear, similar to the real ones in Hermitian systems~\cite{Zhang2023PRB},The effect of NHSE on the WS localization has also been reported in Ref.~\cite{Wang2022PRA}. It will be interesting to ask whether the WS ladder and localization will behave similarly to the real ones in Hermitian systems when the imaginary field becomes amplitude chirped.

To answer the above questions, we introduce the one-dimensional (1D) amplitude-chirped lattices in this work, where the onsite potential is modulated by the function $Fj\cos(2\pi \alpha j)$. Here $F$ is the external field applied to the lattice with a period determined by $\alpha=p/q$ ($p$ and $q$ are co-prime real numbers). Depending on whether $F$ is real or imaginary, we can obtain real or imaginary WS ladders in the eigenenergy spectrum as the field strength increases. The presence of spatially periodic and linearly increasing factors in the onsite potential leads to fascinating phenomena that cannot happen in conventional WS ladders (i.e., the case $\alpha=1$ in our model). In most cases with $q \geq 2$, there are multiple WS ladders with the same or different level spacings, and all the eigenstates become localized in the strong field limit. However, in the lattices with $q=4$, we find that energy-dependent localization phenomenon will emerge, where about half the number of eigenstates are centered around the zero energy at the band center and can extend to a wide region or even the full range of the lattice, even though the field becomes very strong. In addition, in the systems with imaginary field and odd $q$, some of the WS ladders are doubly degenerate in a wide range of field strength, and the corresponding eigenstates are evenly distributed at two neighboring sites instead of one single site. Our work unveils the exotic properties of WS localization in both Hermitian and non-Hermitian amplitude-chirped lattices.

This paper is organized as follows. In Sec.~\ref{Sec2}, we introduce the 1D lattices with amplitude-chirped onsite potential and the corresponding model Hamiltonian. In Secs.~\ref{Sec3} and \ref{Sec4}, we discuss the Hermitian case with real fields and the non-Hermitian case with imaginary fields, respectively. The summary is given in Sec.~\ref{Sec5}.

\section{Model Hamiltonian}\label{Sec2}
We consider the 1D amplitude-chirped lattices described by the following model Hamiltonian
\begin{equation}\label{H1}
	H_1 = \sum_j F j \cos (2\pi \alpha j + \phi) c_j^\dagger c_j + t(c_j^\dagger c_{j+1} + c_{j+1}^\dagger c_j),
\end{equation}
where $c_j^\dagger$ ($c_j$) is the creation (annihilation) operator of spinless fermions at the $j$th site with $V_j=F j \cos (2\pi \alpha j+\phi)$ being the onsite potential. $F$ is the external field applied to the 1D lattice and $\alpha=p/q$ with $p$ and $q$ being co-prime integers determines the period of onsite modulation. $j$ is the site index and $\phi$ is a phase which is set to be $0$ in this work. $t$ is the hopping amplitude between the nearest-neighboring sites, and we will take $t=1$ as the energy unit throughout this paper. If $\alpha=1$, the Hamiltonian reduces to the conventional Stark lattices with a uniform field, where all the eigenstates become localized at a single lattice site when $F$ is strong. If $\alpha<1$, the coexistence of linearly increasing and periodic behaviors in the onsite potential will result in various interesting phenomena in the WS ladder, as will be revealed in the following sections. The size of the lattice is $L$. In this work, we mainly consider the case with $1 \leq j \leq L$ and $\phi=0$. Notice that by choosing different ranges of $j$ or different values of $\phi$, the eigenenergy spectra would be different, but the features of the WS ladder and localization will be the same, see the Appendix.

In Ref.~\cite{Leder2016NatCom}, two such on-site potentials shown in Eq.~(\ref{H1}) with opposite spatial variation are introduced to achieve a zero crossing of the coupling between energy bands. Such lattices can be constructed by using lattice potentials realized in Raman configuration. The positive and negative two-photon detunings are used to realize the spatial chirp of the onsite potential. Our model discussed in this work can also be experimentally realized in similar platforms and we will focus on the WS localization phenomenon in such systems.

In addition to the Hermitian WS ladder with real fields, we also consider the non-Hermitian WS ladder by taking the field imaginary. The Hamiltonian is 
\begin{equation}\label{H2}
	H_2 = \sum_j i F j \cos (2\pi \alpha j + \phi) c_j^\dagger c_j + t(c_j^\dagger c_{j+1} + c_{j+1}^\dagger c_j).
\end{equation}
Notice that the WS ladder in non-Hermitian systems has been investigated in several recent works~\cite{Wang2022PRA,Zhang2023PRB}. The model studied in these works corresponds to our model with $\alpha=1$. By combining the interesting property of non-Hermiticity and the WS ladder, we can expect many more phenomena that cannot happen in the Hermitian cases.

As the field strength increases, the eigenstates in the Stark lattices will become localized. To characterize the localization property of eigenstates, we calculate the inverse participation ratio, which is defined as
\begin{equation}
	IPR = \sum_j \frac{|\psi_{n,j}|^4}{(\left\langle  \psi_n | \psi_n \right\rangle )^2},
\end{equation}
where $\psi_{n,j}$ is the $j$th component of $\psi_n$. For extended states, the $IPR$ value is close to $0$, while for localized states, the $IPR$ is of order $O(1)$.

\section{Hermitian systems with real fields}\label{Sec3}
\begin{figure}[t]
	\includegraphics[width=3.4in]{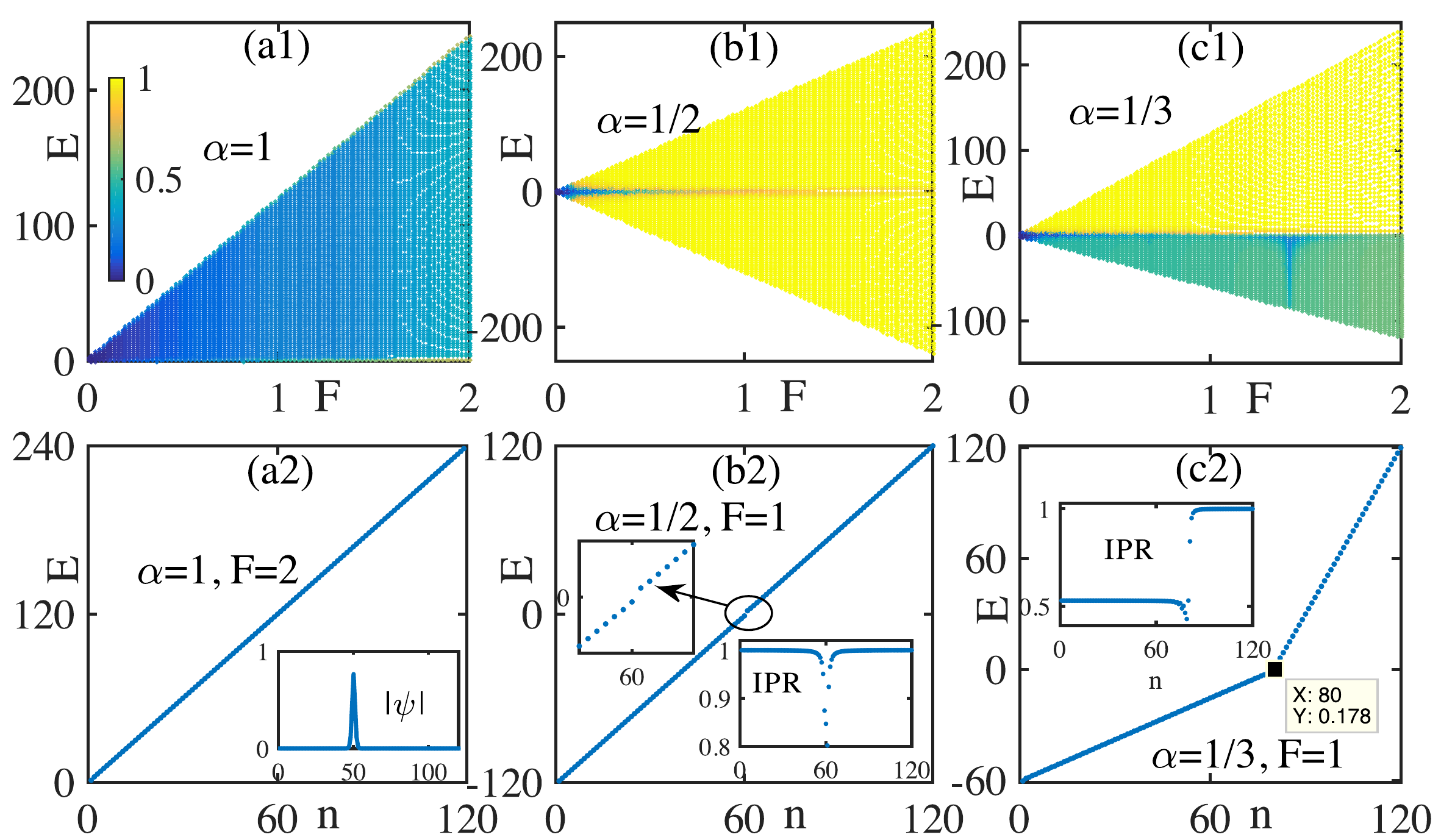}
	\caption{(Color online) Energy spectrum as a function of $F$ for $H_1$ with (a1) $\alpha=1$, (b1) $\alpha=1/2$, and (c1) $\alpha=1/3$. The colorbar indicates the IPR value of the eigenstate. The lower panel shows the corresponding WS ladder with $F$ indicated in the figures. The insets depict the distribution or the IPR values of the eigenstates. The arrow in (b2) indicates the zoom-in of the spectrum around zero energy. The numeric label in (c2) indicates the turning point where the energy changes from negative to positive. Here we set $L=120$ and $\phi=0$.}
	\label{fig1}
\end{figure}

We first check the Hermitian case shown in Eq.~(\ref{H1}), where the system is subjected to a real field. The numerical results given in this work are obtained by diagonalizing the Hamiltonian matrices of the 1D amplitude-chirped lattices under open boundary conditions. If $\alpha=1$, we get the conventional WS ladder in the strong field limit with $E_m=mF$ with $m=1,2,\cdots$. The eigenstate corresponding to the $n$th largest eigenenergy is localized at the $n$th site in the lattice.  In Fig.~\ref{fig1}(a1), we plot the energy spectrum of the system with $\alpha=1$ as a function of $F$. The IPR value indicates that the states are localized when $F$ is large. The WS ladder at $F=2$ is shown in Fig.~\ref{fig1}(a2), where the inset depicts the distribution of one localized eigenstate. Next, we will take $\alpha=p/q$ and check the WS localization in such systems. 

When $\alpha=1/2$, we get $V_j=\pm Fj$ for even and odd $j$s, a staggered onsite potential with the amplitude increased linearly. In the strong field limit, the eigenenergy spectrum is
$E_m = \pm 2m F$ with $m = 1,2,\cdots$. In Fig.~\ref{fig1}(b1), we plot the spectrum as a function of $F$. When the field is strong, there are two WS ladders in the eigenenergy spectrum: one negative and the other positive. The level spacing of both ladders is $\delta E = 2F$. All the eigenstates are localized, and the IPR values are very close to $1$, as shown by the inset in Fig.~\ref{fig1}(b2). Notice that the eigenstates are more localized when compared with the conventional Stark lattice (i.e., $\alpha=1$ in our model). Besides, at $E=0$, which is the band center, the level spacing becomes larger than $2F$. For example, in Fig.~\ref{fig1}(b2), the level spacings between $n=60$ and $61$ levels are larger than $2$ when $F=1$, separating the spectrum into two ladders, as shown by the zoom in around zero energy. The IPR also shows a dip from $1$ to about $0.8$ at the spectral center. Thus in the chirped lattice with $\alpha=1/2$, we can obtain two WS ladders in the eigenenergy spectrum.  

\begin{figure}[t]
	\includegraphics[width=3.4in]{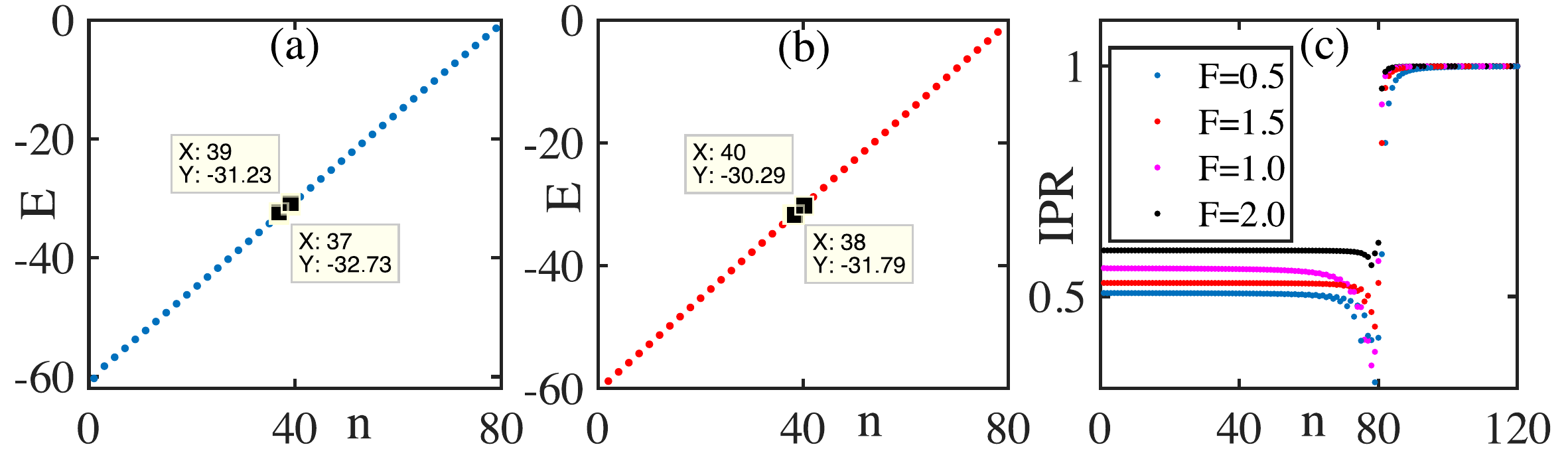}
	\caption{(Color online) (a and b) Two WS ladders with negative energies in the system with $\alpha=1/3$ and $F=1$. The level spacings in these two ladders are the same. (c) Variation of IPR as the field $F$ increases. Other parameters: $L=120$, $\phi=0$.}
	\label{fig2}
\end{figure}

If $\alpha=1/3$, we find three WS ladders in the spectrum when $F$ is strong, as shown in Fig.~\ref{fig1}(c1). From Fig.~\ref{fig1}(c2), we can see that about two-thirds of the eigenenergies are negative. The IPR values for the corresponding eigenstates are constant. By zooming in the spectrum, we can find two WS ladders in this region with the level spacing between the neighboring eigenenergies being $\delta E = 3F/2$, which are respectively labeled by odd and even indices in the spectrum, see Figs.~\ref{fig2}(a) and \ref{fig2}(b). As $F$ increases, these states become more localized as the IPR values gets larger (see Fig.~\ref{fig1}). The other one-third of eigenenergies form another WS ladder with $E>0$ and $\delta E = 3F$, which are more localized than those with $E<0$ since the IPR values jump to $1$. The WS ladders here can be explained as follows. In the strong field limit, the eigenenergy is determined by the diagonal term $V_j=Fj \cos (2j\pi/3)$ in the Hamiltonian, which gives two sorts of eigenvalues: $-Fj/2$ for $mod(j,3)=1$ and $2$, and $Fj$ for $mod(j,3)=0$. The period of cosine function is $3$, so the level spacings in the WS ladders are $3F/2$ and $3F$, respectively. So, for the eigenenergies $E<0$, there are two WS ladders, each with the same level spacing $\delta E = 3F/2$, as shown in Figs.~\ref{fig2}(a) and \ref{fig2}(b). On the other hand, from the perspective of IPR, we can find two plateaus: one plateau with $IPR<1$ for the states with $E<0$, while the other plateau with $IPR=1$ for the $E>0$ states. Besides, the IPR values for the eigenstates with $E<0$ will gradually increase as the field strength $F$ grows, see Fig.~\ref{fig2}(c).  

\begin{figure}[t]
	\includegraphics[width=3.4in]{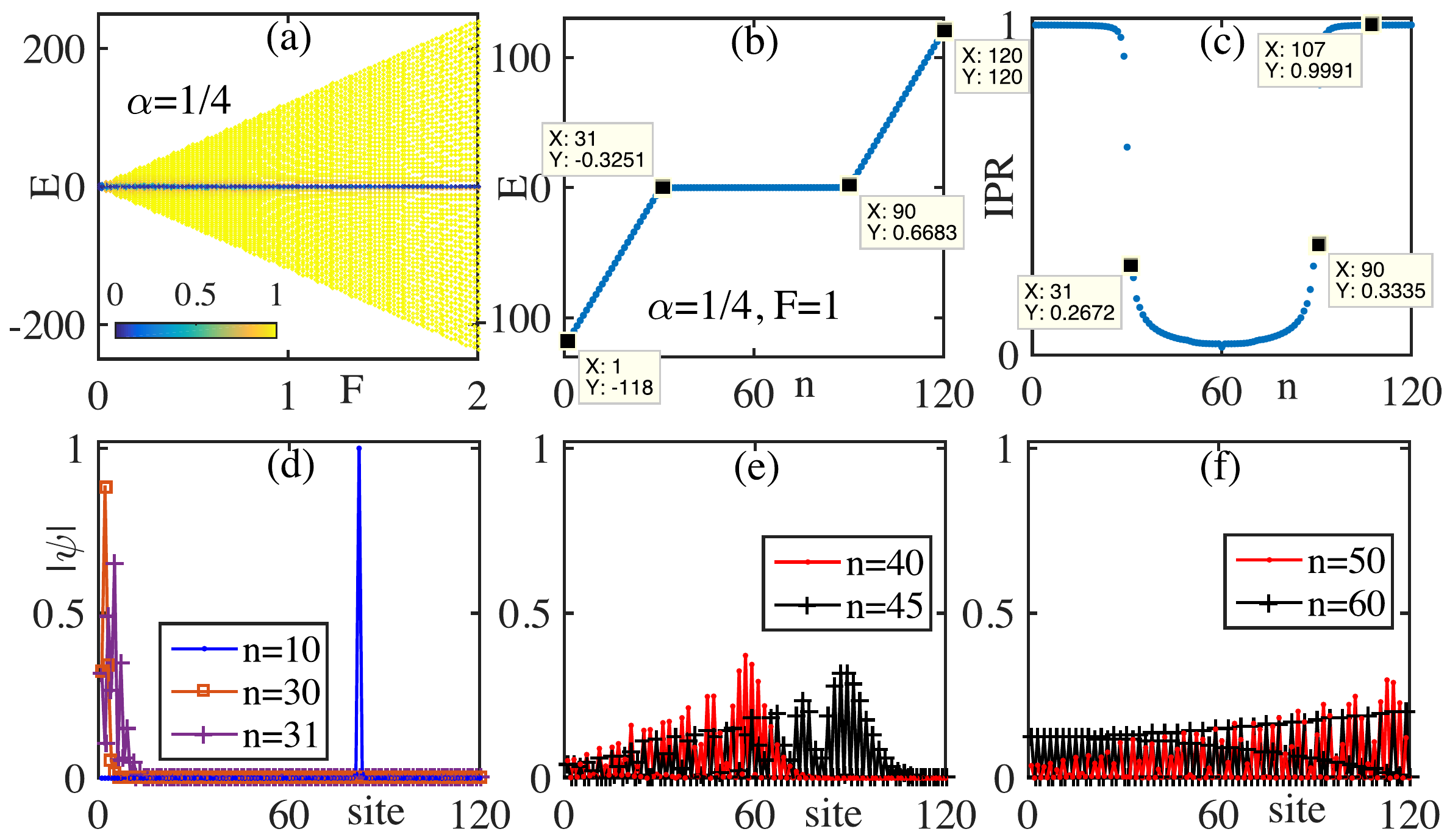}
	\caption{(Color online) (a) Eigenenergy spectrum of the chirped lattice with $\alpha=1/4$. The color bar indicates the IPR values of eigenstates. (b and c) Spectrum and IPR value at $F=1$. (d-f) Distribution of eigenstates. The lattice size is $L=120$ and the phase $\phi=0$.}
	\label{fig3}
\end{figure}

The WS ladders discussed above show different IPR values, meaning that some eigenstates are more localized than others. We may ask whether extended states still exist in such systems even when the field is strong. Thus we further investigate the system with $q=4$. In Fig.~\ref{fig3}(a), we present the energy spectrum of the amplitude-chirped lattice with $\alpha=1/4$. The IPR values indicate that the states with energies away from zero are all localized. We further find that about half the number of eigenenergies is close to the zero energy at the band center. For example, in the lattice with $L=120$, the $31$th to $90$th eigenenergies are gathered around zero, see Fig.~\ref{fig3}(b). The IPR values of the corresponding eigenstates near the band center are much smaller than other states, indicating that they are not localized, as shown in Fig.~\ref{fig3}(c). As the eigenenergy moves closer to zero, the eigenstate extends to broader regions and even to the whole 1D lattice, as exhibited in Figs.~\ref{fig3}(d)-\ref{fig3}(f). In addition, we have two WS ladders with the same level spacing. The central region in the spectrum does not form WS ladders. This is because, in the strong field limit, we have $V_j=Fj \cos(j \pi/2)$, which is zero when $j$ is odd, so there will always be half the number of eigenenergies close to zero when taking the hopping terms into consideration. The remaining eigenenergies, on the other hand, are determined by $V_j=Fj$ or $-Fj$ for even $j$, which correspond to the two ladders in the spectrum. Since the period is $4$, the level spacing of the ladder is $4F$. It is interesting to find that due to the existence of both periodic and linearly increasing behaviors in the onsite potential, the WS localization can be energy-dependent. The states near the band center are extended, while the others are localized. We have also checked the systems with $\alpha=1/6$ and $1/8$ (see the Appendix), but all the states are localized, and no extended states are observed in the strong-field limit, see the Appendix. The reason might be that as $q$ increases, the number of diagonal terms with $V_j=0$ in the Hamiltonian matrix will be smaller and the nonzero terms will take over, leading to the localization of all eigenstates.

\begin{figure}[t]
	\includegraphics[width=3.4in]{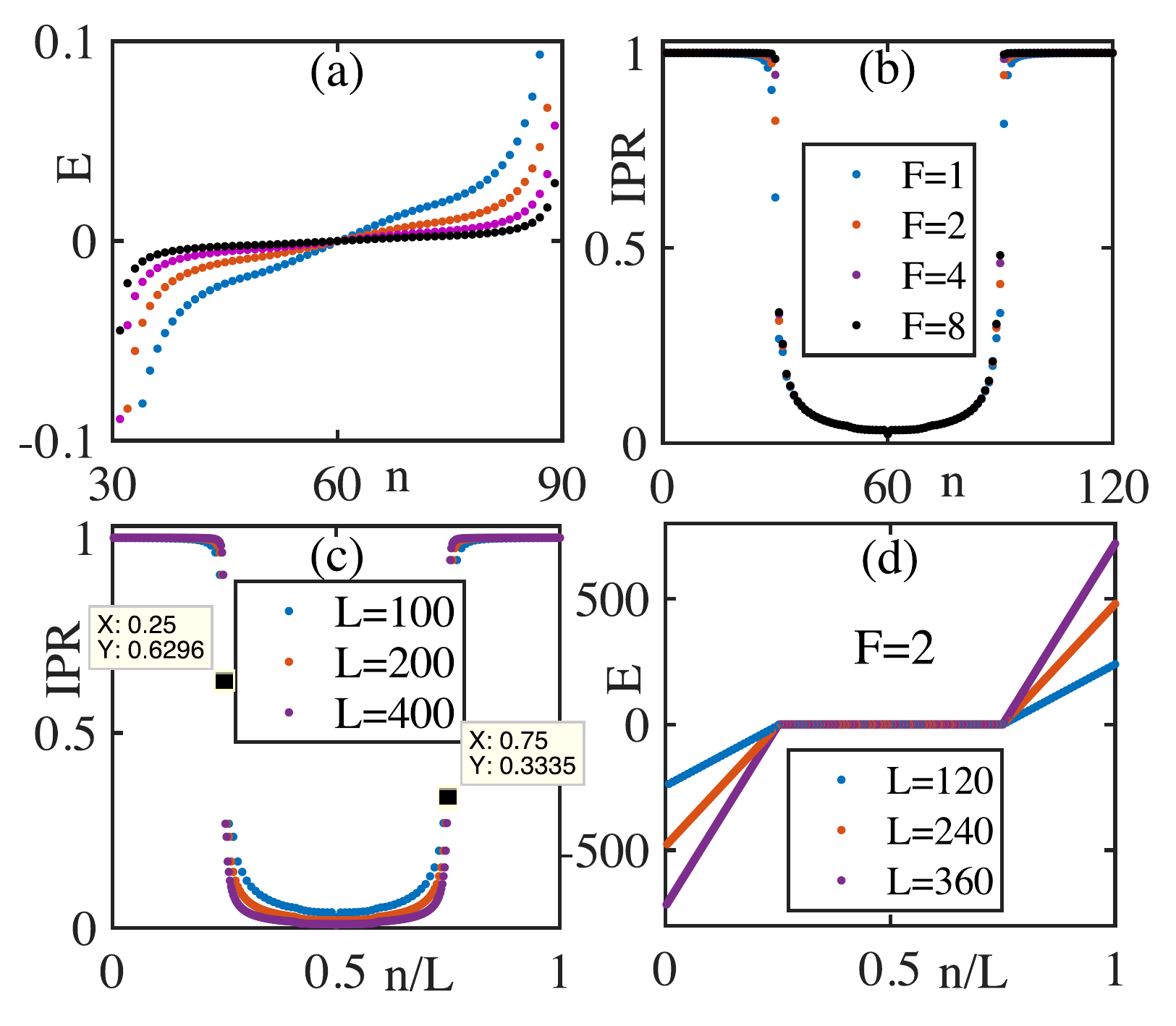}
	\caption{(Color online) Energy spectrum and IPR values for the chirped lattices with $\alpha=1/4$ and different lattice sizes $L$ or field strengths $F$. (a) The spectrum in the middle near zero energy. (b) IPR values under different $F$. Here the lattice size is $L=120$. (c) The change of IPR values of the system with $F=1$ under different sizes. (d) The spectrum at $F=2$ under different sizes. The site number is normalized with respect to the lattice size $L$. Other parameters: $\phi=0$.}
	\label{fig4}
\end{figure}

By increasing $F$, these energies in the middle of the spectrum of the system with $\alpha=1/4$ will move closer to zero, while the IPR values keep almost unchanged even when the field becomes very strong, as shown in Figs.~\ref{fig4}(a) and \ref{fig4}(b). The other states are localized at one site with IPR close to $1$ and form the WS ladder in the spectrum with a constant spacing of $4F$. As the system size increases, the IPR values of the states at the band center (i.e. states with $0.25<n/L<0.75$) will become smaller, indicating that the states are more extended, as shown in Fig.~\ref{fig4}(c). In Fig.~\ref{fig4}(d), we also present the spectra under different lattice sizes and find that the features discussed above remain unchanged. Therefore, the existence of extended states in the chirped lattice with $\alpha=1/4$ is independent of the system size. Similar phenomena also occur in systems with other parameters, such as $\alpha=3/4$. So the energy-dependent localization will always be present in the 1D amplitude-chirped lattices with $q=4$. 

In the Appendix, we provide more numerical results for the 1D amplitude-chirped lattices with different values of $\alpha$ and $\phi$, and combined with the ones we discussed above, we can conclude that the level spacing of the WS ladders in the strong-field limit can be described by the following function:
\begin{equation}
	\delta E = |F q \cos (2\pi k/q + \phi)|,
\end{equation}
where $k$ is an integer chosen from the sequence $(1,2,\cdots,q)$. This is understandable since in the strong field limit, the energy spectrum will be determined by the onsite modulation, which is increases linearly with $j$, and the period of cosine function is $q$.

Recently, single-particle mobility edges have been reported in the 1D mosaic lattices with linearly increasing on-site potential subjected to equally spaced sites~\cite{Dwiputra2022PRB}. The amplitude-chirped lattices we propose here provide another way to obtain energy-dependent localization phenomenon in Stark lattices. The coexistence of spatial period and linearly varying factors in the on-site potential will result in extended states, which can survive in the Stark lattices even in the strong-field limit.

\section{Non-Hermitian systems with imaginary fields}\label{Sec4}
\begin{widetext}
	
	\begin{figure}[t]
		\includegraphics[width=6.5in]{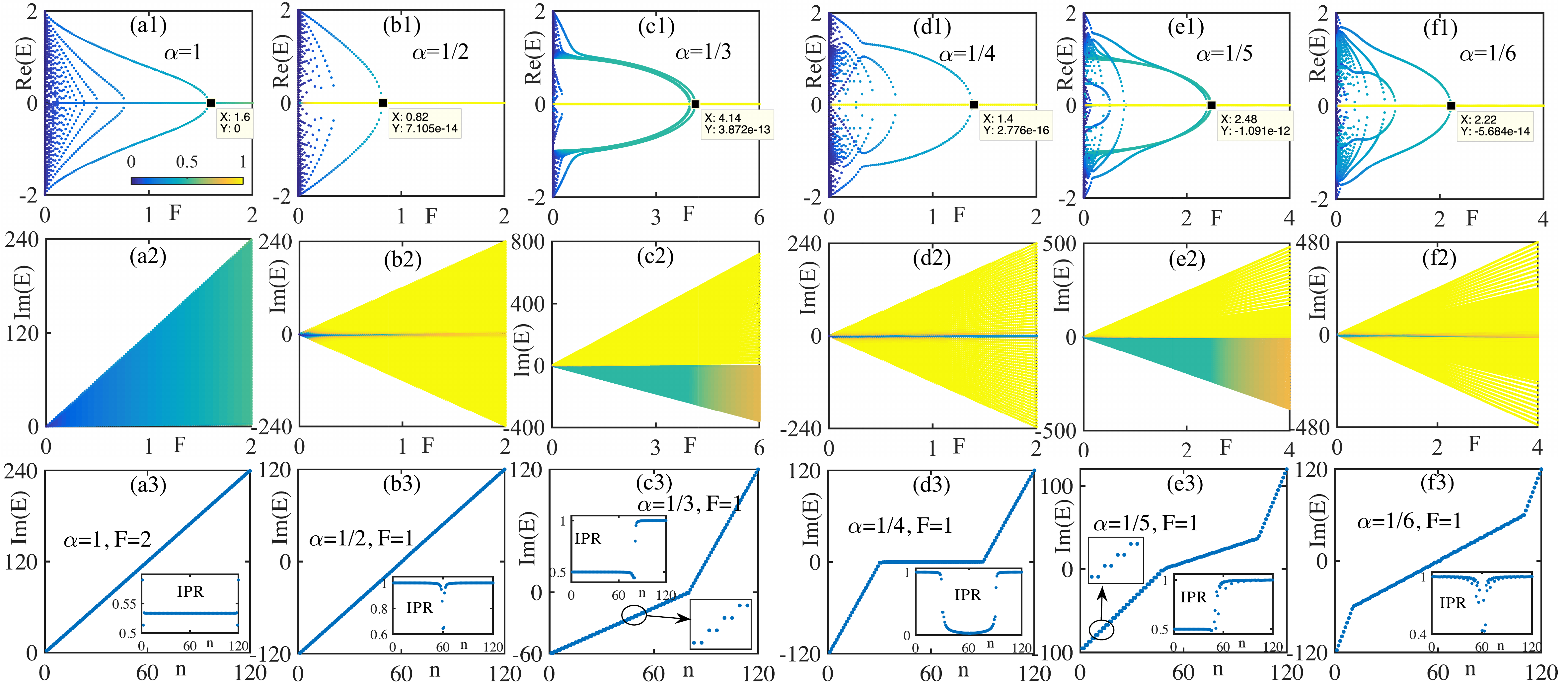}
		\caption{(Color online) Energy spectrum and WS ladders in the non-Hermitian chirped lattices with different $\alpha$. The top and middle panels are the real and imaginary parts of eigenenergy as a function of $F$. The colorbar indicates the IPR value of each eigenstate. The bottom panel presents the imaginary WS ladders under a specific field with the insets being the IPR values for the entire ladder. The eigenenergies are sorted according to the imaginary parts. The lattice size is $L=120$ and $\phi=0$.}
		\label{fig5}
	\end{figure}
	
\end{widetext}

\begin{figure}[t]
	\includegraphics[width=3.4in]{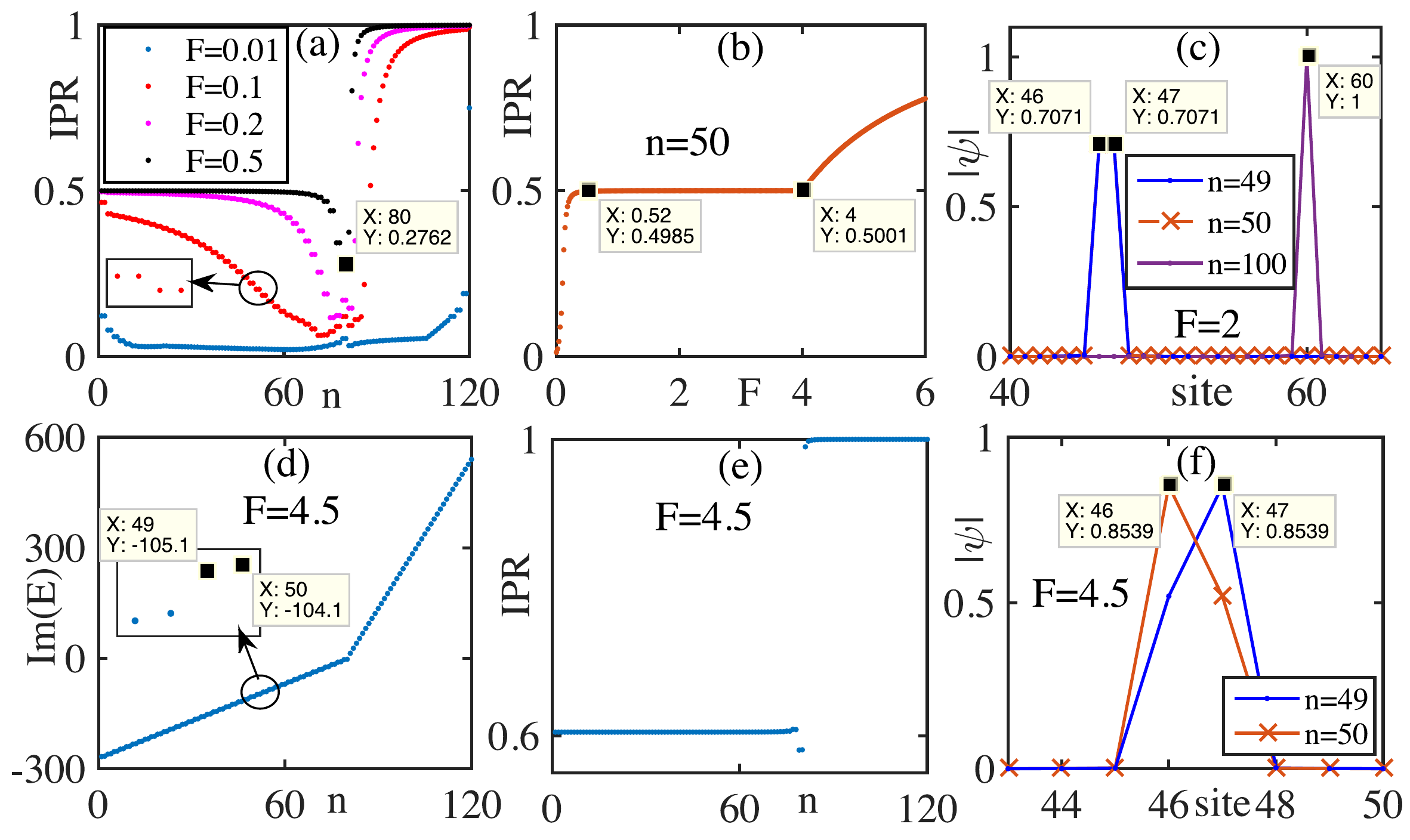}
	\caption{(Color online) IPR values and distribution of eigenstates in the non-Hermitian chirped lattices with $\alpha=1/3$. (a) IPR values under different imaginary fields. (b) The IPR value of the $50$th eigenstate as a function of $F$. (c) The distribution of eigenstates in lattice with $F=2$. (d), (e), and (f) show the spectrum, IPR, and distribution of eigenstates in system with $F=4.5$. Other parameters: $L=120$, $\phi=0$.}
	\label{fig6}
\end{figure}

Now we turn to the non-Hermitian case by replacing $F$ with $iF$ in the Hamiltonian, as shown in Eq.~(\ref{H2}). Most phenomena are the same as in the Hermitian WS ladder discussed in the previous section, except that the energies are imaginary instead of real. However, there are some unique features that do not exist in the Hermitian systems. In the lattice with $\alpha=1$, when the imaginary field is small, the energy spectrum will be continuous in the complex plane, which is reported in Ref.~\cite{Wang2023PRL}. When $F$ becomes stronger, an imaginary WS ladder appears~\cite{Wang2022PRA,Zhang2023PRB}. In the following, we will discuss the WS ladders in the amplitude-chirped lattices with imaginary fields. In Fig.~\ref{fig5}, we present the real and imaginary parts of the eigenenergies in systems with different $\alpha$s. As the field gets strong enough, the spectra will always become purely imaginary. In the bottom panel of Fig.~\ref{fig5}, we present the imaginary WS ladders at the strong-field limit, which behave similarly to the real WS ladders discussed above. One interesting difference happens in lattices with $\alpha=1/3$ and $1/5$. For instance, in the lattice with $\alpha=1/3$ shown in Fig.~\ref{fig5}(c3), it seems that the spectrum is split into two ladders. In fact, about two-thirds of the eigenenergies have negative imaginary parts and they actually form two ladders that are identical, i.e., the ladder is twofold degenerate, as shown by the zoom-in. Moreover, we find that the IPR values for this part locate at $0.5$ and keep unchanged for a wide regime of $F$. To verify this, we calculate the IPR values under different $F$ and $L=120$, which are presented in Fig.~\ref{fig6}(a). When the field is weak, e.g., $F=0.01$, the IPR value is close to zero. As $F$ increases, the IPR value becomes larger and split into two parts, corresponding to the different ladders in the spectrum. The IPR values for the lowest $80$ states are shifted to $0.5$ gradually, while those for the remaining $40$ states are shifted close to $1$. The IPR values show a sharp jump near $n=80$. We further plot the IPR value of the $50$th eigenstate as a function of $F$ and find that it is pinned at $0.5$ in a wide regime with $0.52<F<4$. When $F>4$, the IPR will increase again. This is quite different from the Hermitian case, where the IPR values always increase when the field strength grows, see Fig.~\ref{fig2}(c). The states with $IPR=0.5$ also exhibit a distinctive distribution in the lattice. They are evenly distributed on two neighboring sites, while the states with $IPR=1$ are localized at a single site, as shown in Fig.~\ref{fig6}(c).

The reason behind this phenomenon can be explained as follows. For the states with $IPR=0.5$, the energy is two-fold degenerate. For each eigenenergy, there are two linearly independent states. Suppose that there is one state localized at the $j$th site and denoted by $|j \rangle$, while the other is localized at the $(j+1)$th site and denoted by $|j+1 \rangle$. These two states will hybrid and lead to the following two states:
\begin{equation} 
|\varphi_1 \rangle = \frac{1}{\sqrt{2}} \left( |j \rangle + |j+1 \rangle \right); \quad |\varphi_2 \rangle = \frac{1}{\sqrt{2}} \left( |j \rangle - |j+1 \rangle \right).
\end{equation}
So, the two states distribute on two neighboring sites with equal weight. When $F$ gets stronger than $4$, the twofold degeneracy in the spectrum will be removed, as shown in Fig.~\ref{fig6}(d). The IPR values of the states will not be pinned at $0.5$ but become larger instead. Correspondingly, the eigenstates are not evenly distributed at the two neighboring sites, see Figs.~\ref{fig6}(e) and \ref{fig6}(f). These features also exist in the lattice with $\alpha=1/5$ and other non-Hermitian chirped lattices with odd $q$. So, even though the imaginary WS ladders are similar to the real ones in Hermitian systems, the eigenstates behave differently in certain cases. This reveals the difference between Hermitian and non-Hermitian systems.

\section{Summary}\label{Sec5}
In this work, we have introduced one-dimensional amplitude-chirped lattices, where the WS localization phenomenon is investigated in both Hermitian and non-Hermitian cases. We find that the presence of both spatially periodic and linear growing behaviors in the onsite potential leads to a variety of interesting phenomena that cannot be observed in conventional Stark lattices with only a uniform field. In the strong-field limit, the energy spectrum splits into several WS ladders with different level spacings. Most interestingly, in cases such as $\alpha=1/4$, not all the eigenstates are localized even when the field becomes very strong. Almost half the number of eigenenergies are gathered around the zero energy at the band center, and the corresponding eigenstates can extend over a wide region or even the full range of the lattice, exhibiting the energy-dependent localization phenomenon. We have also studied the WS ladder in the non-Hermitian chirped lattices, where the field is imaginary. The imaginary WS ladders in the energy spectra are similar to the real ones in Hermitian systems, but the ladders can be doubly degenerate and the eigenstates distribute evenly on two neighboring sites in some situations. As to the experimental realization, the amplitude-chirped modulations can be realized in optical lattices using lasers, as shown in Ref.~\cite{Leder2016NatCom}. Our work unveils the exotic properties of WS ladders in amplitude-chirped lattices. The model we studied in this work mainly focus on WS localization under the spatially chirped modulations under open-boundary conditions, and it will be interesting to explore the time evolution of the states in such systems under periodic boundary conditions in future work, similar to the one discussed in the system with constant electric fields~\cite{Niu1989PRB}.

\begin{acknowledgments}
This work is supported by the NSFC (Grant No. 12204326), the Beijing Natural Science Foundation (Grant No. 1232030), the R\&D Program of Beijing Municipal Education Commission (Grant No. KM202210028017), and the Open Research Fund Program of the State Key Laboratory of Low-Dimensional Quantum Physics (Grant No. KF202109).
\end{acknowledgments}

\section*{Appendix}\label{Append}
\newcounter{Sfigure}
\setcounter{figure}{0} 
\renewcommand{\thefigure}{S\arabic{figure}}    
In this Appendix, we give more numerical results of the 1D amplitude-chirped lattices with different parameters. 
\subsection{Spectra with $\alpha=1/5$, $1/6$, $1/7$, and $1/8$}
\begin{widetext}
	
\begin{figure}[t]
	\includegraphics[width=6in]{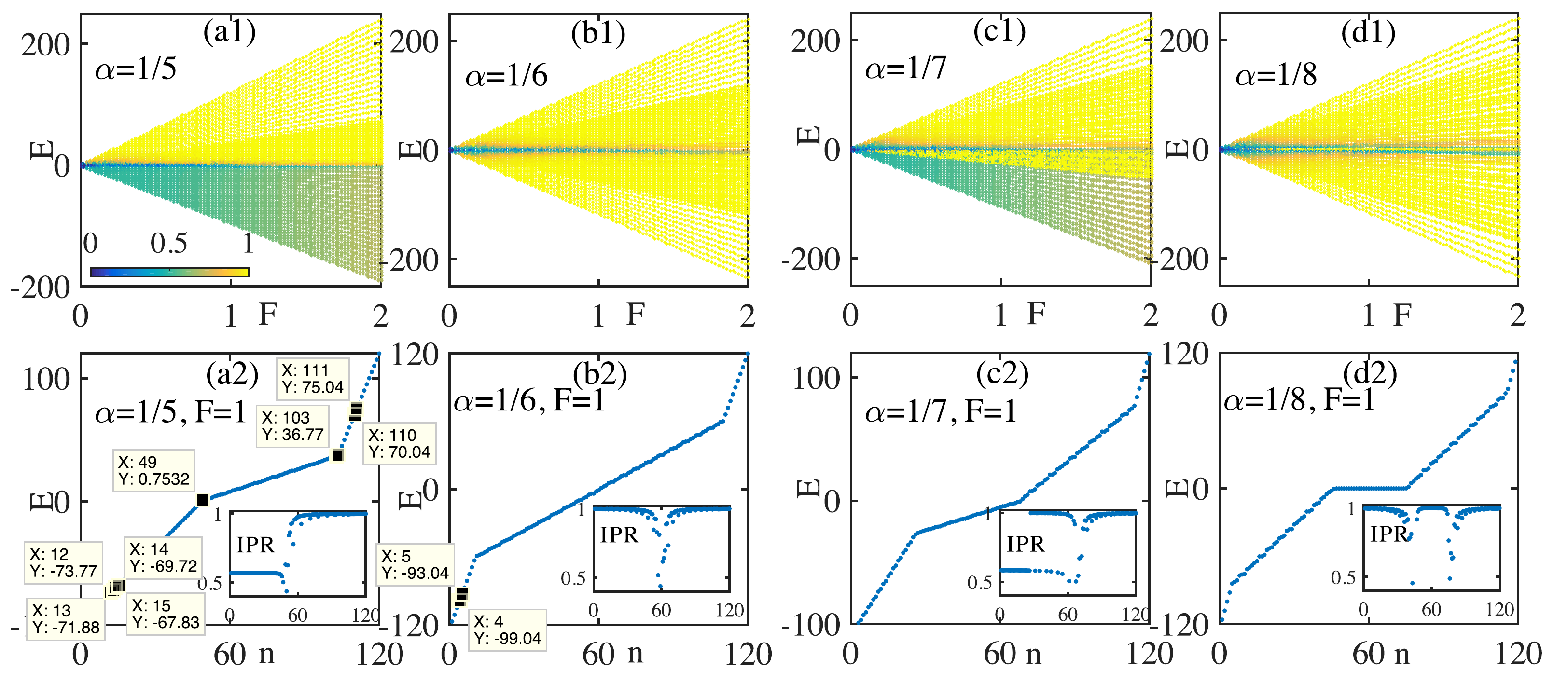}
	\caption{(Color online) (a1-d1) Eigenenergy spectrum as a function of $F$ for systems with $\alpha=1/5$, $1/6$, $1/7$, and $1/8$, respectively. The colorbar indicates the IPR value of the eigenstates. (a2-d2) Corresponding spectra with $F=1$. The insets are the IPR values. Here we take $\phi=0$ and $L=120$.}
	\label{figS1}
\end{figure}

\end{widetext}

In the main text, we have discussed the spectral properties for the 1D Hermitian amplitude-chirped lattices with $\alpha=1/2$, $1/3$, and $1/4$. The results show that due to the coexistence of linearly increasing and periodic behaviors in the modulation, multiple WS ladders will emerge. Here we give more numerical results for such systems with larger $q$ values. In Figs.~\ref{figS1}(a1)-\ref{figS1}(d1), we present the eigenenergy spectrum as a function of the field strength $F$ for systems with $\alpha=1/5$, $1/6$, $1/7$, and $1/8$, respectively. The IPR values of the eigenstates indicate that they are localized in the strong field limit. The corresponding spectra at $F=1$ are shown in Figs.~\ref{figS1}(a2)-\ref{figS1}(d2), where WS ladders can be observed. The spectra split into several parts. The eigenenergies of the parts at band edges form WS ladders while the central part(s) do not, which is similar to the case with $\alpha=1/4$ in the main text. The level spacing of the ladders is closely connected by the value of $\alpha$. For instance, in the spectrum of system with $\alpha=1/5$ and $F=1$ [see Fig.~\ref{figS1}(a2)], we find that the whole spectrum is split into three parts. The first part, i.e., the $1st-49th$ eigenenergies, form two WS ladders, where the energies with odd or even indices have the same level spacing. The numerical result show that the level spacing here is $\delta E =4.05$, which is very close to the value $|5*\cos(4\pi/5)|=4.0451$. The middle part in the spectrum (the $50th-102th$ eigenenergies), however, is not a WS ladder. On the other hand, the remaining part with the $103th-120th$ eigenenergies forms another WS ladder with level spacing being $\delta E=5$. So the level spacing of the WS ladder can always described by a function of the form $\delta E = |F q \cos (2\pi k/q + \phi)|$ with $k\in (1,2,\cdots,q)$, which is determined by the period of modulation. 

\begin{figure}[t]
	\includegraphics[width=3.4in]{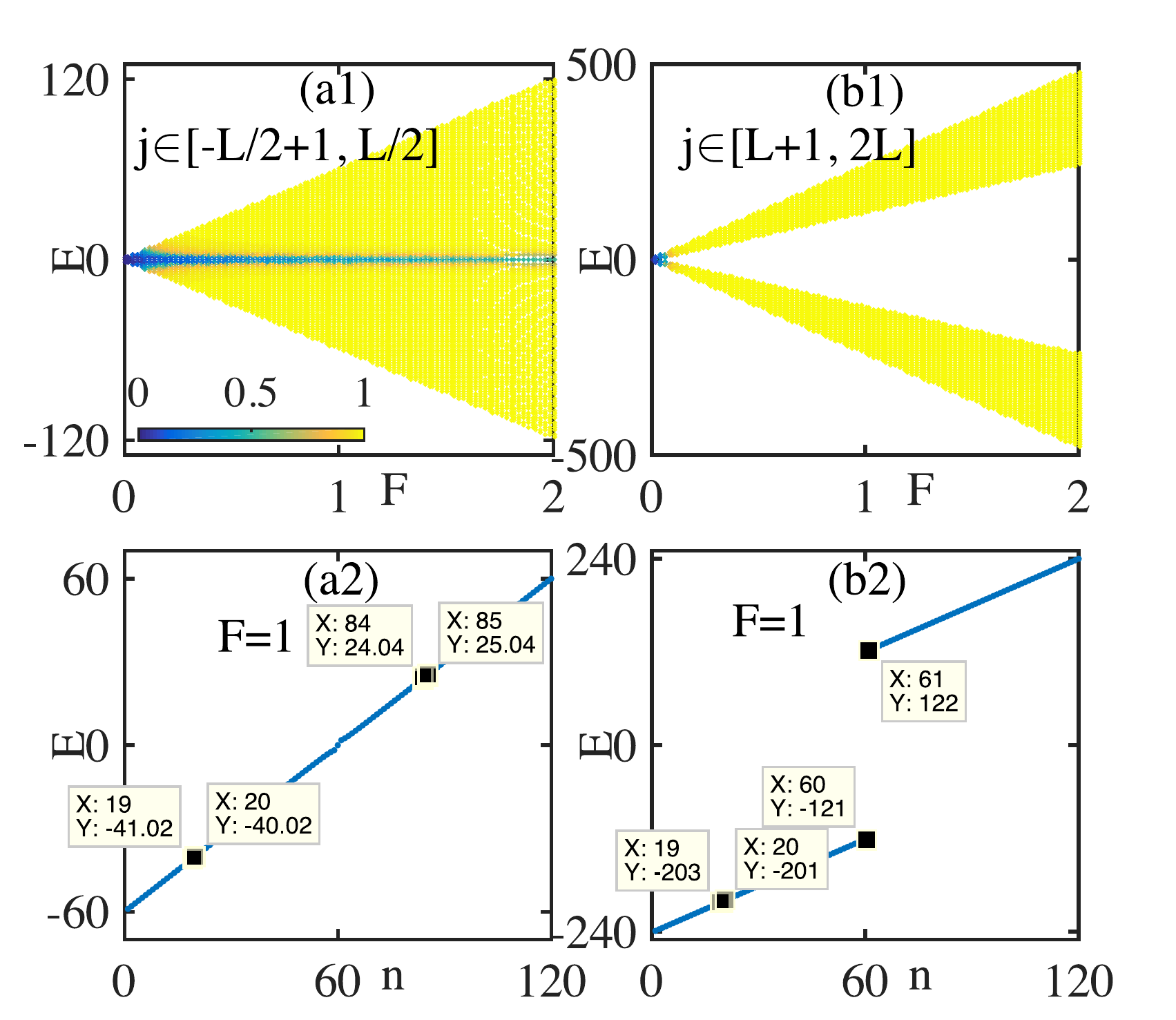}
	\caption{(Color online)  Energy spectra of the 1D chirped lattices with $\alpha=1/2$ and different ranges of $j$. (a1) and (b1) show the spectra as a function of $F$, where the color bar indicates the IPR values of the eigenstates. (a2) and (b2) show the spectra at $F=1$. The lattice size here is $L=120$ and the phase is chosen to be $\phi=0$.}
	\label{figS2}
\end{figure}

\begin{figure}[t]
	\includegraphics[width=3.4in]{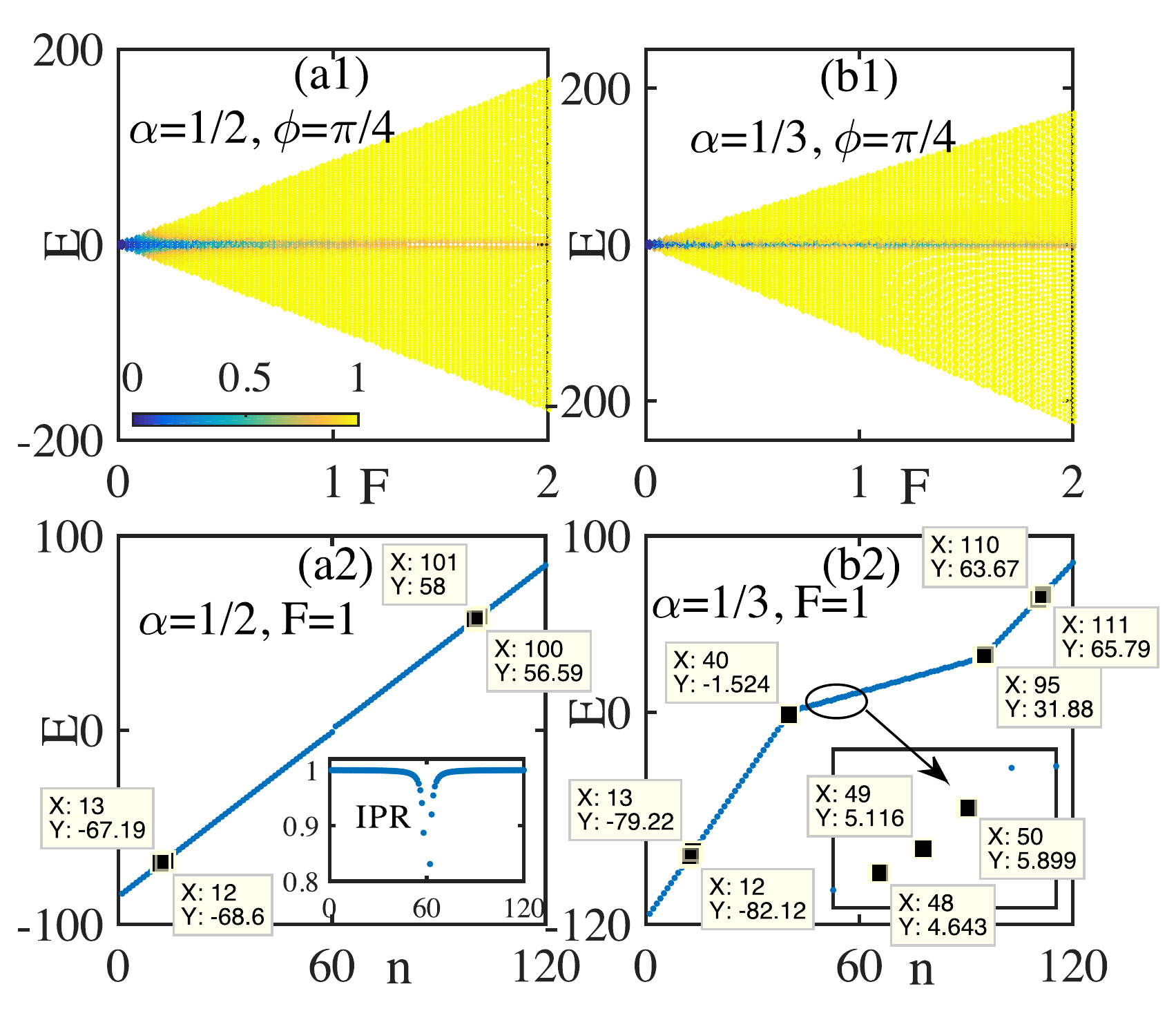}
	\caption{(Color online) Energy spectra of the 1D chirped lattices with $\phi=\pi/4$ when (a) $\alpha=1/2$ and (b) $\alpha=1/3$. The inset in (a2) is the IPR value of the eigenstates of the system with $F=1$. The lattice size is $L=120$ and $j$ is chosen to be $1 \leq j \leq L$.}
	\label{figS3}
\end{figure}

\subsection{System with $j$ in different ranges}
Since the chirped modulation contains both linearly increasing and periodic terms, the energy spectra might depend on the value of $j$. If we change the range of $j$, the linear part in the modulation would be different, which will modify the spectrum accordingly. In Fig.~\ref{figS2}, we present the spectra and WS ladders of the system with $\alpha=1/2$ and two different ranges of $j$: (a) $(-L/2+1) \leq j \leq L/2$ and (b) $(L+1) \leq j \leq 2L$, which are of the same length as the lattices discussed in the main text. For the case with $(-L/2+1) \leq j \leq L/2$, we find that the eigenenergy spectrum as a function of $F$ is very similar to the one with $1 \leq j \leq L$, as shown in Fig.~\ref{figS2}(a1). However, the level spacing of the WS ladder here is $1$ instead of $2$ for the system with $F=1$ [see  Fig.~\ref{figS2}(a2)]. For the amplitude-chirped modulation $Fj \cos (2\pi \alpha j)$, the value of $j$ determines the range of the function. So, when the largest value of $j$ is $L/2$ instead of $L$, the upper and lower limits are also cut in half, and the level spacing of the WS ladder is also halved in the strong field limit.

On the other hand, if we change the range of $j$ to  $(L+1) \leq j \leq 2L$, we find that the spectrum splits into two bands, where the band gap grows almost linearly with the increasing of $F$, as shown in Fig.~\ref{figS2}(b1). This is quit different from the ones we discussed above. The level spacing of the WS ladder is the same as in the system with $1 \leq j \leq L$, see Fig.~\ref{figS2}(b2). So, even though the value range of $j$ will modify the spectra structure, the features of the WS ladders remains the same.

\subsection{System with nonzero $\phi$}
The numerical results presented in the main text are performed for lattices with $\phi=0$, here we consider the case with nonzero phase. In Fig.~\ref{figS3}, we present the energy spectra of the system with $\phi=\pi/4$. For the case with $\alpha=1/2$, we find that the spectrum is very similar to the system with $\phi=0$, only the level spacing of the WS ladder becomes $\delta E=2F\cos(\pi/4)$ with $F=1$, as shown in Fig.~\ref{figS3}(a2). For the case with $\alpha=1/3$, the band structure shown in Fig.~\ref{figS3}(b2) seems quite different from the one with $\phi=0$ in Fig.~\ref{fig1}(c2), where the spectrum split into three parts instead of two. However, the middle part does not form a WS ladder, as indicated by the zooming in. There are still only two WS ladders in the spectrum. Moreover, the level spacing of the ladder with negative energy is around $2.9$, which is close to the value of $|3F \cos(2\pi/3 + \pi/4)|=2.8978$ when $F=1$. The level spacing of the ladder with positive energy is around $2.12$, which  corresponds to the value of $|3F\cos(\pi/4)=2.1213$ with $F=1$. So, the level spacing of the WS ladders is described by $|F q \cos (2\pi k/q + \phi)|$, with $k$ being an integer chosen from the sequence $(1,2,\cdots,q)$.


\begin{thebibliography}{}
\bibitem{Anderson1958}{P. W. Anderson, \href{https://doi.org/10.1103/PhysRev.109.1492}{Phys. Rev. \textbf{109,} 1492 (1958)}.}

\bibitem{Lee1985RMP}{P. A. Lee and T. V. Ramakrishnan, \href{https://doi.org/10.1103/RevModPhys.57.287}{Rev. Mod. Phys. \textbf{57,} 287 (1985)}.}

\bibitem{Evers2008RMP}{F. Evers and A. D. Mirlin, \href{https://doi.org/10.1103/RevModPhys.80.1355}{Rev. Mod. Phys. \textbf{80,} 1355 (2008)}.}

\bibitem{Billy2008Nat}{J. Billy, V. Josse, Z. Zuo, A. Bernard, B. Hambrecht, P. Lugan, D. Clement, L. Sanchez-Palencia, P. Bouyer, and A. Aspect, \href{https://doi.org/10.1038/nature07000}{Nature (London) \textbf{453,} 891 (2008)}.}

\bibitem{Roati2008Nat}{G. Roati, C. D’rrico, L. Fallani, M. Fattori, C. Fort, M. Zaccanti, G. Modugno, M. Modugno, and M. Inguscio, \href{https://doi.org/10.1038/nature07071}{Nature (London) \textbf{453,} 895 (2008)}.}

\bibitem{Luschen2018PRL}{H. P. Lüschen, S. Scherg, T. Kohlert, M. Schreiber, P. Bordia, X. Li, S. Das Sarma, and I. Bloch, \href{https://doi.org/10.1103/PhysRevLett.120.160404}{Phys. Rev. Lett. \textbf{120,} 160404 (2018)}.}

\bibitem{Wiersma1997Nat}{D. S. Wiersma, P. Bartolini, A. Lagendijk, and R. Righini, \href{https://doi.org/10.1038/37757}{Nature (London) \textbf{390,} 671 (1997)}.}

\bibitem{Scheffold1999Nat}{F. Scheffold, R. Lenke, R. Tweer, and G. Maret, \href{https://doi.org/10.1038/18347}{Nature (London) \textbf{398,} 206 (1999)}.}

\bibitem{Schwartz2007Nat}{T. Schwartz, G. Bartal, S. Fishman, and M. Segev, \href{https://doi.org/10.1038/nature05623}{Nature (London) \textbf{446,} 52 (2007)}.}

\bibitem{Dalichaouch1991Nat}{R. Dalichaouch, J. P. Armstrong, S. Schultz, P. M. Platzman, and S. L. McCall, \href{https://doi.org/10.1038/354053a0}{Nature (London) \textbf{354,} 53 (1991)}.}

\bibitem{Pradhan2000PRL}{P. Pradhan and S. Sridhar, \href{https://doi.org/10.1103/PhysRevLett.85.2360}{Phys. Rev. Lett. \textbf{85,} 2360 (2000)}.}

\bibitem{Lahini2009PRL}{Y. Lahini, R. Pugatch, F. Pozzi, M. Sorel, R. Morandotti, N. Davidson, and Y. Silberberg, \href{https://doi.org/10.1103/PhysRevLett.103.013901}{Phys. Rev. Lett. \textbf{103,} 013901 (2009)}.}

\bibitem{Mott1987JPhy}{N. Mott, \href{https://doi.org/10.1088/0022-3719/20/21/008}{J. Phys. C \textbf{20,} 3075 (1987)}.}

\bibitem{Abrahams1979PRL}{E. Abrahams, P. W. Anderson, D. C. Licciardello, and T. V. Ramakrishnan, \href{https://doi.org/10.1103/PhysRevLett.42.673}{Phys. Rev. Lett. \textbf{42,} 673 (1979)}.}

\bibitem{Aubry1980}{S. Aubry and G. André, Ann. Isr. Phys. Soc. \textbf{3,} 133 (1980).}

\bibitem{Wannier1962RMP}{G. H. Wannier, \href{https://doi.org/10.1103/RevModPhys.34.645}{Rev. Mod. Phys. \textbf{34,} 645 (1962)}.}

\bibitem{Fukuyama1973PRB}{H. Fukuyama, R. A. Bari, and H. C. Fogedby, \href{https://doi.org/10.1103/PhysRevB.8.5579}{Phys. Rev. B \textbf{8,} 5579 (1973)}.}

\bibitem{Emin1987PRB}{D. Emin and C. F. Hart, \href{https://doi.org/10.1103/PhysRevB.36.7353}{Phys. Rev. B \textbf{36,} 7353 (1987)}.}

\bibitem{Holthaus1996Phio}{M. Holthaus and D. W. Hone, \href{https://doi.org/10.1080/01418639608240331}{Philos. Mag. B \textbf{74,} 105 (1996)}.}

\bibitem{Hartmann2004NJP}{T. Hartmann, F. Keck, H. Korsch, and S. Mossmann, \href{https://doi.org/10.1088/1367-2630/6/1/002}{New J. Phys. \textbf{6,} 2 (2004)}.}

\bibitem{Schulz2019PRL}{M. Schulz, C. A. Hooley, R. Moessner, and F. Pollmann, \href{https://doi.org/10.1103/PhysRevLett.122.040606}{Phys. Rev. Lett. \textbf{122,} 040606 (2019)}.}

\bibitem{Taylor2020PRB}{S. R. Taylor, M. Schulz, F. Pollmann, and R. Moessner, \href{https://doi.org/10.1103/PhysRevB.102.054206}{Phys. Rev. B \textbf{102,} 054206 (2020)}.}

\bibitem{Guo2021PRL}{Q. Guo, C. Cheng, H. Li, S. Xu, P. Zhang, Z. Wang, C. Song, W. Liu, W. Ren, H. Dong, R. Mondaini, and H. Wang, \href{https://doi.org/10.1103/PhysRevLett.127.240502}{Phys. Rev. Lett. \textbf{127,} 240502 (2021)}.}

\bibitem{Morong2021Nat}{W. Morong, F. Liu, P. Becker, K. S. Collins, L. Feng, A. Kyprianidis, G. Pagano, T. You, A. V. Gorshkov, and C. Monroe, \href{https://doi.org/10.1038/s41586-021-03988-0}{Nature (London) \textbf{599,} 393 (2021)}.}

\bibitem{Leder2016NatCom}{M. Leder, C. Grossert, L. Sitta, M. Genske, A. Rosch, and M. Weitz,  \href{https://doi.org/10.1038/ncomms13112}{Nat. Commun. \textbf{7,} 13112 (2016)}. }

\bibitem{Cao2015RMP}{H. Cao and J. Wiersig, \href{https://doi.org/10.1103/RevModPhys.87.61}{Rev. Mod. Phys. \textbf{87,} 61 (2015).}}

\bibitem{Konotop2016RMP}{V. V. Konotop, J. Yang, and D. A. Zezyulin, \href{https://doi.org/10.1103/RevModPhys.88.035002}{Rev. Mod. Phys. \textbf{88,} 035002 (2016).}}

\bibitem{Ganainy2018NatPhy}{R. El-Ganainy, K. G. Makris, M. Khajavikhan, Z. H. Musslimani, S. Rotter, and D. N. Christodoulides, \href{https://doi.org/10.1038/nphys4323}{Nat. Phys. \textbf{14,} 11 (2018).}}

\bibitem{Ashida2020AiP}{Y. Ashida, Z. Gong, and M. Ueda, \href{https://doi.org/10.1080/00018732.2021.1876991}{Advances in Physics \textbf{69,} 249 (2020).}}

\bibitem{Bergholtz2021RMP}{E. J. Bergholtz, J. C. Budich, and F. K. Kunst, \href{https://doi.org/10.1103/RevModPhys.93.015005}{Rev. Mod. Phys. \textbf{93,} 015005 (2021).}}	

\bibitem{Bender1998PRL}{C. M. Bender and S. Boettcher, \href{https://doi.org/10.1103/PhysRevLett.80.5243}{Phys. Rev. Lett. \textbf{80,} 5243 (1998).}}

\bibitem{Bender2002PRL}{C. M. Bender, D. C. Brody, and H. F. Jones, \href{https://doi.org/10.1103/PhysRevLett.89.270401}{Phys. Rev. Lett. \textbf{89,} 270401 (2002).}}

\bibitem{Bender2007RPP}{C. M. Bender, \href{https://doi.org/10.1088/0034-4885/70/6/R03}{Rep. Prog. Phys. \textbf{70,} 947 (2007).}}

\bibitem{Yao2018PRL1}{S. Yao and Z. Wang, \href{https://doi.org/10.1103/PhysRevLett.121.086803}{Phys. Rev. Lett. \textbf{121,} 086803 (2018).}}

\bibitem{Yao2018PRL2}{S. Yao, F. Song, and Z. Wang, \href{https://doi.org/10.1103/PhysRevLett.121.136802}{Phys. Rev. Lett. \textbf{121,} 136802 (2018).}}

\bibitem{Gong2018PRX}{Z. Gong, Y. Ashida, K. Kawabata, K. Takasan, S. Higashikawa, and M. Ueda, \href{https://doi.org/10.1103/PhysRevX.8.031079}{Phys. Rev. X \textbf{8,} 031079 (2018).}}

\bibitem{Shnerb1998PRL}{N. M. Shnerb and D. R. Nelson, \href{https://doi.org/10.1103/PhysRevLett.80.5172}{Phys. Rev. Lett. \textbf{80,} 5172 (1998).}}

\bibitem{Jiang2019PRB}{H. Jiang, L.-J. Lang, C. Yang, S.-L. Zhu, and S. Chen, \href{https://doi.org/10.1103/PhysRevB.100.054301}{Phys. Rev. B \textbf{100,} 054301 (2019).}}

\bibitem{Zeng2020PRR}{Q.-B. Zeng and Y. Xu, \href{https://doi.org/10.1103/PhysRevResearch.2.033052}{Phys. Rev. Research \textbf{2,} 033052 (2020).}}

\bibitem{Liu2021PRB1}{Y. Liu, Y. Wang, X. J. Liu, Q. Zhou, and S. Chen, \href{https://doi.org/10.1103/PhysRevB.103.014203}{Phys. Rev. B \textbf{103,} 014203 (2021).}}

\bibitem{Liu2021PRB2}{Y. Liu, Q. Zhou, and S. Chen, \href{https://doi.org/10.1103/PhysRevB.104.024201}{Phys. Rev. B \textbf{104,} 024201 (2021).}}

\bibitem{Wang2023PRL}{Q. Wang, C. Zhu, X. Zheng, H. Xue, B. Zhang, and Y. D. Chong, \href{https://doi.org/10.1103/PhysRevLett.130.103602}{Phys. Rev. Lett. \textbf{130,} 103602 (2023)}.}

\bibitem{Zhang2023PRB}{Y. Zhang and S. Chen, \href{https://doi.org/10.1103/PhysRevB.107.224306}{Phys. Rev. B \textbf{107,} 224306 (2023)}.}

\bibitem{Wang2022PRA}{H.-Y. Wang and W.-M. Liu, \href{https://doi.org/10.1103/PhysRevA.106.052216}{Phys. Rev. A \textbf{106,} 052216 (2022)}.}

\bibitem{Dwiputra2022PRB}{D. Dwiputra and F. P. Zen, \href{https://doi.org/10.1103/PhysRevB.105.L081110}{Phys. Rev. B \textbf{105,} L081110 (2022)}.}

\bibitem{Niu1989PRB}{Q. Niu, \href{https://doi.org/10.1103/PhysRevB.40.3625}{Phys. Rev. B \textbf{40,} 3625 (1989)}.}

\end{thebibliography}
\end{document}